\documentclass[12pt]{JHEP3}

\usepackage{graphicx}
\usepackage{psfrag}

\def\beq{\begin{equation}}
\def\eeq{\end{equation}}
\def\bea{\begin{eqnarray}}
\def\eea{\end{eqnarray}}
\def\nn{\nonumber}
\def\nl{\nonumber\\}
\def\bra#1{\langle #1|}
\def\ket#1{| #1\rangle}
\def\roughly#1{\mathrel{\raise.3ex\hbox
{$#1$\kern-.75em\lower1ex\hbox{$\sim$}}}}

\def\sss{\scriptscriptstyle}

\def\bd{B_d^0}

\def\bs{B_s^0}
\def\bsbar{{\overline{B_s^0}}}

\def\btos{{\bar b} \to {\bar s}}

\def\ANPu{{\cal A}^u}

\def\mw{M_{\sss W}}
\def\gf{G_{\sss F}}
\def\lft{{\sss L}}
\def\rht{{\sss R}}
\def\bskk{\bs\to K^+ K^-}
\def\bdpipi{\bd\to \pi^+ \pi^-}
\def\wt{\widetilde}

\newcommand{\av}[1]{\langle #1 \rangle}

\title{Supersymmetric Contributions to {\boldmath $\bskk$}}

\author{Seungwon Baek\\
Laboratoire Ren\'e J.-A. L\'evesque, Universit\'e de Montr\'eal, \\
C.P. 6128, succ. centre-ville, Montr\'eal, QC, Canada H3C 3J7 \\
and \\
Department of Physics, Yonsei University, Seoul 120-749, Korea \\
E-mail: \email{swbaek@kias.re.kr}}

\author{David London\\
Laboratoire Ren\'e J.-A. L\'evesque, Universit\'e de Montr\'eal, \\
C.P. 6128, succ. centre-ville, Montr\'eal, QC, Canada H3C 3J7 \\
E-mail: \email{london@lps.umontreal.ca}}

\author{Joaquim Matias\\
IFAE, Universitat Aut\`onoma de Barcelona, \\
08193 Bellaterra, Barcelona, Spain \\
E-mail: \email{matias@ifae.es}}

\author{Javier Virto\\
IFAE, Universitat Aut\`onoma de Barcelona, \\
08193 Bellaterra, Barcelona, Spain \\
E-mail: \email{jvirto@ifae.es}}

\abstract{Inspired by the existing calculation of $B\to\pi K$ decays
in supersymmetry (SUSY), we evaluate the dominant SUSY contributions
to $\bskk$. We show that the observables of this process can be
significantly modified in the presence of SUSY. In particular, the
branching ratio can be increased considerably compared to the
prediction of the standard model (SM). The effect is even more
dramatic for the CP-violating asymmetries $A_{dir}$ and
$A_{mix}$. These asymmetries, expected to be small in the SM
($A_{dir}$ is predicted to take only positive values), change
drastically with SUSY contributions.  The measurement of these
observables can therefore be used to detect the presence of physics
beyond the SM, and put constraints on its parameters.}


\preprint{UdeM-GPP-TH-05-142  \\ UAB-FT-594 }

\begin{document}

\section{Introduction}

Current measurements of $B$ decays show hints of physics beyond the
standard model (SM), in CP-violating asymmetries in penguin-dominated
$\btos q{\bar q}$ transitions ($q=u,d,s$) \cite{sin2betapeng}, in
triple-product asymmetries in $B \to \phi K^*$\cite{BVVTP,phiKstarTP},
in the polarization measurements of $B \to \phi
K^*$\cite{BphiKstar_exp,BphiKstarNP, BphiKstarSM} and $B\to \rho K^*$
\cite{BrhoKstar_exp,BrhoKstar}, and in $B\to\pi K$ decays (branching
ratios and CP asymmetries) \cite{BKpiexp,BKpidecays0,
BKpidecays1,BKpidecays2,BKpidecays2bis,BKpidecays3,BKpidecays4}.
These discrepancies are (almost) all not yet statistically
significant, being in the 1--2$\sigma$ range. However, if these hints
are taken together, the statistical significance
increases. Furthermore, they are intriguing since they all point to
new physics (NP) in $\btos$ transitions. For this reason it is
interesting to consider the effect of NP on $B$ decays dominated by
the quark-level $\btos$ process.

One such decay is $\bskk$. In the SM, its amplitude is given
approximately by
\beq
A(\bskk) = - P' - T' ~.
\label{BsKKampSM}
\eeq
Here the prime on the amplitude stands for a strangeness-changing
decay. In the above, $P'$ and $T'$ are the gluonic penguin amplitude
and the color-favored tree amplitude, respectively. These are
estimated to obey the hierarchy $P':T' \sim 1:{\bar\lambda}$, where
${\bar\lambda} \sim 0.2$ \cite{GHLR}. There are other diagrams, but
they are expected to be $O({\bar\lambda}^2)$, and have been neglected
above.

The amplitude $P'$ is actually composed of three pieces, $P'_u$,
$P'_c$ and $P'_t$, where the subscript refers to the internal quark in
the loop:
\bea
P' & = & V_{ub}^* V_{us} \, P'_u + V_{cb}^* V_{cs} \, P'_c + V_{tb}^*
V_{ts} \, P'_t \nn\\
&\simeq & V_{cb}^* V_{cs} \, (P'_c - P'_t) ~.
\eea
In writing the second line, we have used the unitarity of the
Cabibbo-Kobayashi-Maskawa (CKM) matrix to eliminate the $V_{tb}^*
V_{ts}$ term, and we have dropped the $V_{ub}^* V_{us}$ term since
$|V_{ub}^* V_{us}| \ll |V_{cb}^* V_{cs}|$.

Note: even though the term $V_{ub}^* V_{us} \, (P'_u - P'_t)$ is at
the level of other terms we have neglected, it can be retained by
redefining the $T'$ amplitude:
\beq
T' \to T' + V_{ub}^* V_{us} \, (P'_u - P'_t) ~.
\label{Tprimeamp}
\eeq
In the rest of the paper we will adopt this redefinition. Thus, $T'$
has both a tree and a (small) penguin component.

At the quark level, $\bskk$ is described by $\btos u{\bar u}$. There
are many potential NP contributions, which at the quark level take the
form $\bra{K^+ K^-} {\bar b} \Gamma_i s \, {\bar u} \Gamma_j u
\ket{\bs}$, where the $\Gamma_{i,j}$ represent Lorentz structures, and
color indices are suppressed. (We expect the size of all NP
contributions to be at most of the order of $|P'|$.) This picture can
be simplified by considering the strong phases.

In Ref.~\cite{DL}, it was observed that the NP strong phases are
negligible compared to that of the (dominant) SM contribution $P'$.
(Note: each NP contribution can in principle have a different strong
phase.) Briefly, the argument goes as follows. All strong phases are
due to rescattering from intermediate states, with a suppression
factor of about 10--20. In the SM, the $P'_c$ strong phase arises
principally from the rescattering of the ${\bar b} \to {\bar c} c
{\bar s}$ tree diagram, $T'_c$. Since $T'_c$ is about 10--20 times
bigger than $P'_c$, a strong phase of $O(1)$ is generated. By
contrast, the NP strong phases can arise only from
``self-rescattering,'' i.e.\ rescattering from NP operators
themselves. As a consequence, these phases are only about 5--10\% as
large as that of $P'$, and are therefore negligible. This leads to a
great simplification: if one neglects the NP strong phases, one can
combine all NP matrix elements into a single NP amplitude, with a
single weak phase:
\beq
\sum c_{ij} \bra{K^+ K^-} {\bar b} \Gamma_i s \, {\bar u} \Gamma_j u
\ket{\bs} \equiv \ANPu e^{i \Phi_u} ~,
\eeq
where the $c_{ij}$ are the coefficients of the operators and
$\Phi_u$ is the effective NP weak phase.

Note that while this argument --- that the NP strong phases are
negligible --- is quite general, there are still ways of evading this
result. This can occur, for example, if certain NP amplitudes are
larger than $|P'|$ and do not contribute to $\bskk$, but still
contribute to the rescattering. This situation is perhaps unlikely,
but the reader should be aware of these caveats.

Note also that the $T'$ strong phase is expected to be small. Thus,
the relative strong phase between $T'$ and the NP is small compared to
that of $P'$. Below, we will take this to be $(0 \pm 10)^\circ$.

In a previous article, three of us (DL, JM, JV) showed that one can
measure the parameters $|\ANPu|$ and $\Phi_u$ by combining
measurements of $\bskk$ and $\bdpipi$ \cite{LMV}. In the present
paper, we consider the generation of $|\ANPu|$ and $\Phi_u$ within a
specific NP model: minimal supersymmetry (SUSY).

Naively, one would guess that all NP contributions to $|\ANPu|$ and
$\Phi_u$ are suppressed by $\mw^2/M_{\sss NP}^2$, where $M_{\sss NP}
\sim 1$ TeV, and are therefore small. However, there are SUSY
contributions involving squark-gluino loops. Since these involve the
strong coupling constant $\alpha_s$, they are proportional to
$\alpha_s/M_{\sss NP}^2$, and so can compete with the SM contributions
which are of order $\alpha/\mw^2$ [$(\alpha_s/\alpha)(\mw^2/M_{\sss
NP}^2) \sim 1$]. Thus, there are large SUSY contributions to the NP
parameters. Indeed, these are the dominant effects, and are the only
ones which are considered below.  As we will see, one can generate an
$|\ANPu|$ of the same order as $|P'|$, so that the amplitude for
$\bskk$ can be written
\beq
A(\bskk) = - P' - T' + \ANPu e^{i \Phi_u} ~.
\eeq

The effect of SUSY on the $\bskk$ observables can therefore be
sizeable, and we examine it here. We begin in Sec.~2 by establishing
the SM predictions for the various observables in $\bskk$. In Sec.~3,
we evaluate the SUSY contributions to the NP parameters $|\ANPu|$ and
$\Phi_u$. With this information, in Sec.~4 we calculate the combined
effect of the SM and SUSY on the $\bskk$ observables. We note that the
presence of SUSY can dramatically change the values of these
observables. Thus, their measurements can both establish the presence
of NP and constrain the SUSY parameter space. We conclude in Sec.~5.

\section{\boldmath $\bskk$: SM Results} \label{secSM}

We begin with general definitions of CP-violating asymmetries. For the
decay $\bs\to f$, where $f$ is a CP eigenstate, one can measure two
such asymmetries. The direct CP asymmetry takes the form
\beq
A_{dir} = { \left\vert {\cal A} \right\vert^2 - \left\vert {\bar{\cal
A}} \right\vert^2 \over \left\vert {\cal A} \right\vert^2 + \left\vert
{\bar{\cal A}} \right\vert^2} \label{Adir}~,
\eeq
where ${\cal A}$ is the amplitude for $\bs\to f$. ${\bar{\cal A}}$ is
formed from ${\cal A}$ by changing the sign of the weak phases. The
mixing-induced (indirect) CP asymmetry takes the form
\beq
A_{mix} = -2\ { {\rm Im} \left( e^{-i\phi_s} {\cal A}^* {\bar{\cal A}}
\right) \over \left\vert {\cal A} \right\vert^2 + \left\vert
{\bar{\cal A}} \right\vert^2} ~,
\label{eq:CPS}
\eeq
where $\phi_s$ is the phase of $\bs$--$\bsbar$ mixing.

We now turn to specific expectations for $\bskk$ within the SM.  This
process has three observables: the two CP asymmetries mentioned above,
and the branching ratio. Without calculation, we can estimate the
expected size of the CP asymmetries. For $\bskk$, since the amplitude
$T'$ is subdominant, to leading order this decay is described by a
single amplitude, $V_{cb}^* V_{cs} \, (P'_c - P'_t)$. As such, in the
SM the direct CP asymmetry is expected to be small, of order $|T'/P'|
\sim {\bar\lambda} \sim 20\%$. Similarly, the mixing-induced CP
asymmetry approximately measures $\phi_s$. Since $\phi_s$ is also
expected to be very small (in the Wolfenstein parametrization
\cite{Wolfenstein}, ${\rm Im} V_{ts} \sim 5\%$), this asymmetry is
expected to be correspondingly small.

In order to calculate the SM predictions for these three
observables, we need the magnitudes and relative weak and strong
phases of $P'$ and $T'$.  The relative weak phase is $\gamma$, one
of the three interior CP-violating angles of the unitarity
triangle. This phase can be obtained from a fit to a variety of
other measurements, some non-CP-violating. The latest analysis
gives $\gamma = {61^{+7}_{-5}}^\circ$ \cite{CKMfitter}. Note that
this error includes uncertainties in theoretical quantities. This
value will be used in our analysis.

For the magnitudes and relative strong phase of $P'$ and $T'$, we can
proceed in one of two ways. One approach is to use a particular
theoretical framework to calculate these quantities (see, for
instance, \cite{BKpidecays1,Chen:2001sx}). Alternatively, one can use
measurements of $\bdpipi$, along with flavor SU(3) symmetry, to obtain
$P'$ and $T'$ \cite{BKpidecays2,BKpidecays3,Fleischer:1999,LonMat}. In
this paper, we adopt the latter approach.

Neglecting small contributions, the amplitude for the decay $\bdpipi$
can be written
\beq
A(\bdpipi) = - P - T ~.
\eeq
As above, we can write
\bea
P & = & V_{ub}^* V_{ud} \, P_u + V_{cb}^* V_{cd} \, P_c + V_{tb}^*
V_{td} \, P_t \nn\\
&= & V_{ub}^* V_{ud} \, (P_u - P_t) + V_{cb}^* V_{cd} \, (P_c - P_t)
~.
\eea
The difference compared to $P'$ is that one cannot neglect the first
term. On the other hand, we can absorb it into the definition of $T$:
\beq
T \to T + V_{ub}^* V_{ud} \, (P_u - P_t) ~.
\eeq
Thus, $T$ is not a pure tree amplitude, but contains a penguin
amplitude.

As with $\bskk$, there are three measurements involving $\bdpipi$: the
two CP asymmetries and the branching ratio. These suffice to determine
the magnitudes and relative strong phase of $P$ and $T$, given the
knowledge of $\gamma$. Using flavor SU(3) symmetry, these can be
related to the magnitudes and relative strong phase of $P'$ and $T'$
\cite{BKpidecays2,BKpidecays3,Fleischer:1999,LonMat}:
\bea
\left\vert {T' \over T} \right\vert & = & \left\vert {V_{us}
  \over V_{ud}} \right\vert \cal{R_C} ~, \nn\\
\left\vert {P'/T' \over P/T} \right\vert & = & \left\vert {V_{cs}
  V_{ud} \over V_{cd} V_{us}} \right\vert \xi ~.
\eea
In the SU(3) limit we have
${\cal R_C} = 1$, $\xi = 1$ and $\theta' = \theta$, where
$\theta'$ and $\theta$ are the relative strong phases of $P'$ and
$T'$, and $P$ and $T$, respectively.

Due to U-spin breaking, ${\cal R_C}$ gets both factorizable and
non-factorizable contributions. The former have recently been
calculated using QCD sum rules \cite{Mannel} and found to be sizeable:
\beq {\cal R_C} = 1.76^{+0.15}_{-0.17} ~. \label{Rc} \eeq
It should be noticed, however, that factorizable corrections
are
absent in the double ratio $(P'/T')/(P/T)$.  In our analysis, we use
the central value of ${\cal R_C}$. For the other quantities, we take
$\xi = 1.0 \pm 0.2$ (which we vary), and $\theta' - \theta =
0^\circ$. Whenever we refer to the U-spin limit, we will mean $\xi =
1$ and $\theta' = \theta$, but always taking the value of
Eq.~(\ref{Rc}) for ${\cal R_C}$.

\begin{figure}
\begin{center} \psfrag{BR}{${\scriptstyle
BR^{\sss SM}_{\sss KK}\, (\times 10^6)}$} \psfrag{Adir}{${\scriptstyle A_{dir\ \sss KK}^{\sss SM}}$}
\psfrag{Amix}{${\scriptstyle A_{mix\ \sss KK}^{\sss SM}}$}
\psfrag{cc}{${\scriptstyle \!\!\!\!\!---\ \xi=1.1}$}
\psfrag{aa}{${\scriptstyle
\!\!\!\!\!\cdot\cdot\cdot\cdot\cdot\cdot\cdot\cdot\ \xi=0.9}$}
\psfrag{bb}{${\scriptstyle \!\!\!\!\!-\!-\!-\!-\ \xi=1}$}
\includegraphics[height=5cm,width=15.2cm]{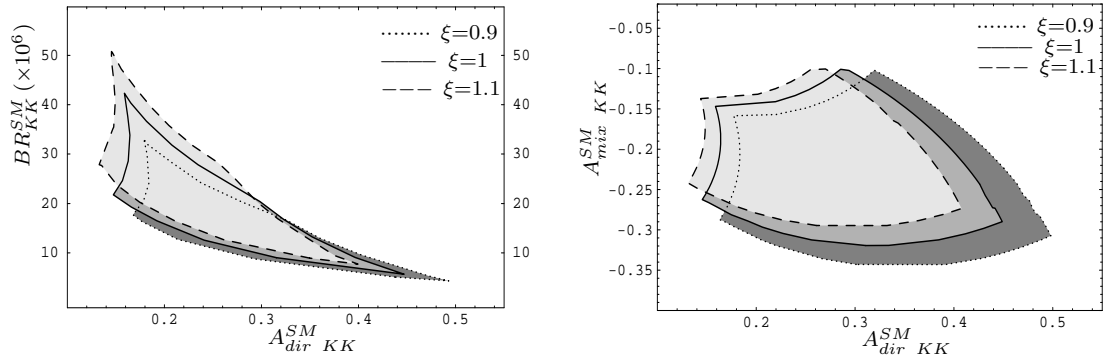}
\end{center}
\vspace{-.5cm}
\caption{\small Correlations between the observables
$A_{dir\ \sss KK}^{\sss SM}-BR^{\sss SM}_{\sss KK}$ and $A_{dir\ \sss KK}^{\sss SM}-A_{mix\
\sss KK}^{\sss SM}$, for $\gamma=61^\circ$ and $\xi=1$, $\xi=0.9$ and
$\xi=1.1$.}
\label{CorrSM}
\end{figure}

With the experimental measurements of $\bdpipi$ and the theoretical
values for the SU(3)-breaking parameters, we can obtain $P'$ and $T'$,
which allow us to compute the SM expectations for the $\bskk$
observables. The latest $\bdpipi$ data is:
$$
\begin{array}{rcl}

BR(B_d^0\to \pi^+\pi^-)&=&\left\{ \begin{array}{ll}
(5.5\pm 0.5)\times 10^{-6} & \mathtt{BaBar}\ \cite{Aubert:2005ni}\\
(4.4\pm 0.7)\times 10^{-6} & \mathtt{Belle}\ \cite{Chao:2003ue}\\
(5.0\pm 0.4)\times 10^{-6} & \mathtt{Average} \end{array}
\right. \\

&&\\

A_{dir}(B_d^0\to \pi^+\pi^-)&=&\left\{
\begin{array}{ll}
-0.09\pm 0.16 & \mathtt{BaBar}\ \cite{Aubert:2005av}\\
-0.52\pm 0.14 & \mathtt{Belle}\ \cite{AbeLP05}\\
-0.33\pm 0.11 & \mathtt{Average}
\end{array}
\right. \\

&&\\

A_{mix}(B_d^0\to \pi^+\pi^-)&=&\left\{
\begin{array}{ll}
0.30\pm 0.17 & \mathtt{BaBar}\ \cite{Aubert:2005av}\\
0.67\pm 0.17 & \mathtt{Belle}\ \cite{Abe:2005dz}\\
0.49\pm 0.12 & \mathtt{Average}
\end{array} \right. \end{array}
$$\\
Regarding $BR_{\sss KK}^{\sss SM}$, it is sometimes
more useful to present the ratio of branching ratios of $\bs\to
K^+K^-$ and $\bd\to \pi^+\pi^-$: $R_{d}^{s} \equiv BR(\bs \to
K^+ K^-)/BR(\bd \to \pi^+ \pi^-)$ \cite{LonMat}.  The SM $\bskk$
predictions for all four quantities are shown in Table 1 (see also
\cite{BKpidecays3}).
Obviously
these values are correlated. Fig.~\ref{CorrSM} illustrates the main
correlations between the observables, for different values of the SU(3)
breaking parameter $\xi$.

\begin{table} \label{tabb1}
\begin{center}
\begin{tabular}{|c|c|c|c|c|}
\hline
 & $BR_{\sss KK}^{\sss SM}\ (\times 10^6)$ & $R_{d}^{s \; SM}$ &
$A_{dir\ \sss KK}^{\sss SM}$ & $A_{mix\ \sss KK}^{\sss SM}$ \\
\hline
$\begin{array}{c}
\gamma=61^\circ\\
\xi=1 \end{array}$
& $(6.4,42.6)$ & $(1.2,9.3)$ & $(0.15,0.45)$ & $(-0.32,-0.10)$ \\
\hline
$\begin{array}{c}
\gamma=61^\circ\\ \xi=1\pm 0.2
\end{array}$
& $(4.2,61.9)$ & $(0.8,13.5)$ & $(0.12,0.56)$ & $(-0.38,-0.09)$ \\
\hline
$\begin{array}{c}
\gamma=(61^{+7}_{-5})^\circ\\ \xi=1
\end{array}$
& $(5.0,60.7)$ & $(0.9,13.2)$ & $(0.08,0.58)$ & $(-0.34,0.08)$ \\
\hline
\end{tabular}
\end{center}
\caption{\small SM predictions for the branching ratio and mixing
induced and direct CP-asymmetries. The impact of the uncertainty in the
U-spin breaking parameter $\xi$ and CKM-angle $\gamma$ is shown.}
\end{table}
\vskip0.3truecm

Note that the CP asymmetries are allowed to take large values. This
does not imply that our above argument about the expected smallness of
these asymmetries is incorrect. Rather, it points to the largeness of
the present experimental errors.

Despite the large regions, there is still room for NP. If any of the
correlations is found to lie outside of the allowed regions, this is a
signal of physics beyond the SM.

\section{\boldmath SUSY Contributions to $|\ANPu|$ and $\Phi_u$}

In this section we evaluate the SUSY contributions to $|\ANPu|$ and
$\Phi_u$. We adopt the following procedure:
\begin{enumerate}

\item We consider all operators generated at the heavy scale, taken to
be $\mw$. We compute the SUSY contributions to the coefficients of
these operators.

\item Using the renormalization group, we run the operator
coefficients down to $m_b$. Operator mixing is included here.

\item We compute the matrix elements of the various operators at
$m_b$. This allows us to calculate $|\ANPu|$ and $\Phi_u$.

\end{enumerate}
We closely follow the approach of Grossman, Neubert and Kagan (GNK)
\cite{trojan}. One difference is that GNK are interested in
isospin-violating effects (``trojan penguins''), while we consider
both isospin-conserving and isospin-violating contributions to
$|\ANPu|$ and $\Phi_u$. Another difference is that GNK calculate the
NP contributions to $B\to\pi K$, while we concentrate on $\bskk$. Here
the quark-level calculation is the same, and so our computation can be
considered as a check.

We begin by listing all the operators which are generated by the new
physics at the heavy scale.  The NP effective hamiltonian is
\cite{trojan}
\beq
H_{\rm eff}^{\sss NP}=\frac{\gf}{\sqrt{2}} \left[
\sum_{i,q=u,d} \left( c_i^q(\mu) \, O_i^q + \tilde{c}_i^q(\mu) \,
\tilde{O}_i^q \right) + C_{8g}(\mu) \, Q_{8g} + {\tilde
C}_{8g}(\mu) \, {\tilde Q}_{8g} \right] ~,
\label{H}
\eeq
where
\beq
\begin{array}{ll}
O_1^q=(\bar{b}_\alpha s_\alpha)_{V-A}(\bar{q}_\beta
q_\beta)_{V+A}\qquad ,  & O_2^q=(\bar{b}_\alpha
s_\beta)_{V-A}(\bar{q}_\beta
q_\alpha)_{V+A} ~, \\
O_3^q=(\bar{b}_\alpha s_\alpha)_{V-A}(\bar{q}_\beta
q_\beta)_{V-A}\qquad , & O_4^q=(\bar{b}_\alpha
s_\beta)_{V-A}(\bar{q}_\beta
q_\alpha)_{V-A} ~, \\
O_5^q=(\bar{b}_\alpha q_\alpha)_{V-A}(\bar{q}_\beta
s_\beta)_{V+A}\qquad , & O_6^q=(\bar{b}_\alpha
q_\beta)_{V-A}(\bar{q}_\beta s_\alpha)_{V+A} ~, \\
Q_{8g} = (g_s / 8 \pi^2) m_b {\bar b} \sigma_{\mu \nu} (1 -
\gamma_5)G^{\mu \nu} s ~. &
\label{ops}
\end{array}
\label{NPops}
\eeq\\
In the above, $\alpha$ and $\beta$ are color indices, and the
subscript $V\pm A$ indicates that the Lorentz structure between quarks
is $\gamma^\mu (1 \pm \gamma_5)$. Despite the fact that, at the quark
level, $\bskk$ is $\btos u{\bar u}$, $d$-quark operators must be
included above since they mix with the $u$-quark operators upon
renormalization to $m_b$. Note that the above list includes the
chromomagnetic operator $Q_{8g}$. The operators $\tilde{O}_i^q$ and
${\tilde Q}_{8g}$ are obtained from $O_i^q$ and $Q_{8g}$ by chirality
flipping.

The above list of operators includes new-physics contributions to
electroweak-penguin operators. As we will see, these effects can be
significant. This shows that, although the SM electroweak-penguin
contributions to $\bskk$ are negligible, the same does not hold for
the NP.

We now must compute the NP contributions to the Wilson coefficients
$c_i^q(\mu)$ and $\tilde{c}_i^q(\mu)$ at the heavy scale $\mu = \mw$.
As discussed above, the dominant contribution comes from QCD penguin
and box diagrams with squark-gluino loops.

For our analysis, we follow Ref.~\cite{trojan} and take the general
minimal supersymmetric standard model at the electroweak scale without
assuming any flavor models at high energies (e.g.\ at the scale of
grand unification). Here, the SUSY flavor-changing neutral current
problem is avoided by assuming that the down squark is decoupled from
the strange and bottom squarks \cite{trojan}. That is, we write
\bea
{\tilde d}_\lft & = & {\tilde d}_\lft^0 \nn\\
{\tilde s}_\lft & = & \cos\theta_\lft {\tilde s}_\lft^0 -
\sin\theta_\lft e^{-i \delta_\lft} {\tilde b}_\lft^0 \nn\\
{\tilde b}_\lft & = & \sin\theta_\lft e^{i \delta_\lft} {\tilde
s}_\lft^0 + \cos\theta_\lft {\tilde b}_\lft^0 ~.
\eea
In the above, the superscript `0' indicates gauge eigenstates, and
$\delta_\lft$ is a new CP-violating phase.  There are similar
expressions for the right-handed squarks. The $\left\vert
\theta_{\lft,\rht} \right\vert$ are taken to be $\le 45^\circ$.
Similarly, the up squark is assumed to be decoupled from the charm
squark, and up-top squark mixing can also be ignored \cite{trojan}.
With these approximations, the Wilson coefficients $c_{5,6}^q(\mu)$
and $\tilde{c}_{5,6}^q(\mu)$ vanish; the others are given in Appendix
A.

Once we have calculated, for given values of the SUSY parameters, the
Wilson coefficients at $M_W$, the next step is to compute the
renormalization-group running of these, including operator mixing,
down to $m_b$. The details of the computation are given in Appendix B.

The final step in the program is to compute the hadronic matrix
elements of the operators in Eq.~(\ref{NPops}) for $\bskk$. These are
calculated using the naive factorization approach.

We define
\beq
A_X^Y\equiv
\bra{K^-}(\bar{b}u)_X\ket{B_s^0}\bra{K^+}(\bar{u}s)_Y\ket{0} ~,
\eeq
where $X$ and $Y$ refer to Lorentz structures. The pseudoscalar nature
of the mesons implies that
\beq
\begin{array}{l}
A_{V+A}^{V+A}=A_{V-A}^{V+A}=-A_{V+A}^{V-A}=-A_{V-A}^{V-A}\equiv
A\\ \scriptscriptstyle{ } ~, \\
A_{S+P}^{S+P}=A_{S-P}^{S+P}=-A_{S+P}^{S-P}=-A_{S-P}^{S-P}\equiv S ~,
\end{array}
\eeq
which define the hadronic quantities $A$ and $S$. After Fierz
rearranging and factorization, the matrix elements of the operators
read:
\beq
\begin{array}{ll}
\av{O_1^u}=2\eta S \qquad  , & \av{O_2^u}=2S ~, \\
\av{O_3^u}=-\eta A \qquad  , & \av{O_4^u}=-A ~, \\
\av{O_5^u}=A \qquad , & \av{O_6^u}=\eta A ~,
\label{matrixops}
\end{array} \eeq
where $\eta=1/N_C=1/3$. The matrix elements of the operators
$\tilde{O_i^u}$ are just $\av{\tilde{O}_i^u}=-\av{O_i^u}$, the minus
sign coming from the change $A \to -A$ and $S \to -S$. Finally, we
define
\beq
\chi\equiv -2S/A;\qquad \bar{c}_i^q\equiv c_i^q-\tilde{c}_i^q
\eeq

The NP amplitude can now be written as
\begin{eqnarray}
\bra{K^+K^-} H_{\rm eff}^{\sss NP} \ket{\bs} & = &
\frac{\gf}{\sqrt{2}} \left[
-\chi(\frac{1}{3}\bar{c}_1^u+\bar{c}_2^u)-\frac{1}{3}(\bar{c}_3^u-\bar{c}_6^u)-
(\bar{c}_4^u-\bar{c}_5^u) \right. \nn\\
& & \hskip10truemm \left. -\lambda_t \frac{2\alpha_s}{3\pi}
\bar{C}_{8g}^{\rm eff}\Big( 1+\frac{\chi}{3}\Big) \right] A ~,
\label{KKmatrixelement}
\end{eqnarray}
where the coefficients $\bar{c}_i^u$ are evaluated at $m_b$. The
hadronic quantities $\chi$ and $A$ can be calculated in terms of the
meson masses, form factors and decay constants. Using the following
expressions for the factorized amplitudes,
\beq
\begin{array}{ll}
\bra{K^+}\bar{u}\gamma_\mu\gamma_5s\ket{0}=i\sqrt{2}f_Kp_\mu
 & \displaystyle
\bra{K^+}\bar{u}\gamma_5s\ket{0}=-\frac{i\sqrt{2}f_Km_K^2}{m_u+m_s}\\
\displaystyle \bra{K^-}\bar{b}\gamma^\mu
u\ket{B_s^0}=\frac{m_B^2-m_K^2}{q_-^2}q_-^\mu F^{B\to
K}\hspace{1cm} & \displaystyle \bra{K^-}\bar{b}
u\ket{B_s^0}=\frac{1}{m_b}(m_B^2-m_K^2) F^{B\to K}~, \end{array}
\eeq
with $q_-^\mu\equiv q_B^\mu-q_{K^-}^\mu=q_{K^+}^\mu\equiv p^\mu$, we
find that
\begin{eqnarray}
\label{chi}
\chi&=&\frac{2m_K^2}{m_b(m_u+m_s)}\simeq 1.18 ~, \\
&&\nn\\
A&=& i\sqrt{2}(m_B^2-m_K^2)f_KF^{B\to K}\simeq i\,1.37 \,{\rm
GeV}^3 ~.
\end{eqnarray}
In the above, the values of the masses, decay constants and form
factors are taken from Refs.~\cite{BKpidecays1,pdg}.

\section{\boldmath $\bskk$: SM $+$ SUSY}

We are now ready to calculate the values of the various $\bskk$
observables in the presence of SUSY. We begin with $BR_{\sss KK}$ and
$A_{dir\ \sss KK}$.  The parameters $P'$ and $T'$ are taken from the
SM analysis (Sec.~2). The NP parameters $|\ANPu|$ and $\Phi_u$ are the
modulus and argument of the amplitude in Eq.~(\ref{KKmatrixelement}).
Finally, we must address the question of the relative strong phase of
$T'$ and $\ANPu$. If the factor $(P'_u - P'_t)$ were not present in
$T'$ [Eq.~(\ref{Tprimeamp})], we would say that the strong phase of
$T'$ is the same as that of the NP, i.e.\ it is negligible and
$\delta_{\sss T'} - \delta_{\sss NP} = 0$. However, $(P'_u - P'_t)$
{\it is} present. And since $P'_u$ can have a non-negligible strong
phase due to rescattering from the $\btos u {\bar u}$ tree diagram,
the relative strong phase of $T'$ and $\ANPu$ can be nonzero. We take
$\delta_{\sss T'} - \delta_{\sss NP} = (0 \pm 10)^\circ$. The
quantities $BR_{\sss KK}$ and $A_{dir\ \sss KK}$ can now be obtained.

The mixing-induced CP asymmetry, $A_{mix\ \sss KK}$, can also be
affected by the presence of SUSY. However, in order to compute the
allowed range, we must take into account the fact that this NP will
also affect the $\bs$--$\bsbar$ mixing angle $\phi_s$. The SM predicts
$\phi_s \approx 0$, because the combination of CKM matrix elements
$(V_{ts}^* V_{tb})^2$ is real to a very good approximation
\cite{Wolfenstein}. On the other hand, in the SUSY scenario we
consider, sizeable $\phi_s$ is possible. Barring the simultaneous
existence of $LL$ and $RR$ mixing, we get the following expression for
$\phi_s$:
\bea
 \phi_s &=& \arg\Bigg[
1 +
 e^{-2i \delta_L}
 \;{\sin^22 \theta_L \over \lambda_t^2}
 \;{\alpha_s^2 \over \alpha_W^2}
 \;{m_W^2 \over m_{\wt{g}}^2}
 \;{1 \over S_0(x_t)} \nl
    &\times& \Bigg|\frac{11}{18} \Big(
    G(x_{\tilde b_L\tilde g},x_{\tilde b_L\tilde g})
    + G(x_{\tilde s_L\tilde g},x_{\tilde s_L\tilde g})
    - 2 G(x_{\tilde b_L\tilde g},x_{\tilde s_L\tilde g}) \Big)
    \nonumber\\
   &&\mbox{}- \frac{2}{9} \Big(
    F(x_{\tilde b_L\tilde g},x_{\tilde b_L\tilde g})
    + F(x_{\tilde s_L\tilde g},x_{\tilde s_L\tilde g})
    - 2 F(x_{\tilde b_L\tilde g},x_{\tilde s_L\tilde g})\Big) \Bigg|
    \Bigg] \,.
\eea
Here $x_t\equiv m_t^2/m_W^2$ and $\lambda_t\equiv V_{tb} V_{ts}^*$.
The loop functions $F$ and $G$, are given in Appendix A, and $S_0$ is
\beq
S_0(x)=\frac{x^4-12x^3+15x^2-4x+6x^3\ln{x}}{4(x-1)^3}
\eeq
For the case of $RR$-mixing, we can use the same formula with $L
\leftrightarrow R$. This allows us to compute $A_{mix\ \sss KK}$ in
the presence of SUSY.

The complete expressions for $BR_{\sss KK}$, $A_{dir\ \sss KK}$ and
$A_{mix\ \sss KK}$ depend on a number of unknown SUSY
parameters. These are the gluino and squark masses, and the angles
$\theta_{\sss L,R}$ and $\delta_{\lft, \rht}$. (The relative strong
phase $\delta_{\sss T'} - \delta_{\sss NP}$ has been discussed above.)
Our aim here is to see how the space of allowed values for the $\bskk$
observables is increased with respect to that of the SM alone
(Sec.~\ref{secSM}). For this purpose, we take the following
ranges/values for the SUSY parameters. For the angles, we take $-\pi/4
\le \theta_{\lft, \rht} \le \pi/4$ and $-\pi \le \delta_{\lft, \rht}
\le \pi$. For the masses we take $m_{\tilde{g}} = m_{\tilde{d}_{\lft,
\rht}} = m_{\tilde{b}_{\lft, \rht}} = 250 \, {\rm GeV}$, $250~{\rm
GeV} \le m_{\tilde{u}_{\lft, \rht}} \le 1000~{\rm GeV}$, and $500~{\rm
GeV} \le m_{\tilde{s}_{\lft, \rht} } \le 1000~{\rm GeV}$. We also take
$m_{\tilde{q}_\rht} = m_{\tilde{q}_\lft}$.

Some further constraints are imposed on the set of input parameters.
First, the same SUSY contributions to $\bskk$ will also affect $B\to\pi K$
decays. In particular, there will be effects on the quantities $R_*$
and $A_{\rm CP}(\pi^+K^0)$ \cite{trojan}, whose definitions and values are
\cite{HFAG}
\bea
R_* &\equiv& \frac{BR(B^+\rightarrow \pi^+K^0)+BR(B^-\rightarrow \pi^-\bar{K}^0)}
{2[BR(B^+\rightarrow \pi^0K^+)+BR(B^-\rightarrow \pi^0K^-)]}=1.00\pm 0.08 \label{R*}\\
A_{\sss CP}(\pi^+K^0)&\equiv& \frac{BR(B^+\rightarrow \pi^+K^0)-BR(B^-\rightarrow \pi^-\bar{K}^0)}
{BR(B^+\rightarrow \pi^+K^0)+BR(B^-\rightarrow \pi^-\bar{K}^0)}=-0.020\pm 0.034\qquad
\eea
In order to incorporate these two constraints we follow the approach
in Ref.~\cite{trojan} where QCD-factorization is used, except for the
strong phase related to $A_{\rm CP}$ which we take as a free
parameter. Second, there are bounds from $BR(B\to X_s+\gamma)$ and
$\Delta m_s$ \cite{Alexander:2005cx}:\footnote{We take a wider range
for $BR(B\to X_s+\gamma)$ to allow for various theoretical
uncertainties. }
\bea
&&2.92\times 10^{-4}<BR(B\to X_s+\gamma)<4.12\times 10^{-4} \label{BRXsgamma}\\
&&\Delta m_s/\Delta m_s^{SM}>0.9797 \label{Deltams} \eea
The measured values (\ref{R*})-(\ref{Deltams}) will therefore put
additional constraints on the SUSY parameter space, which are
taken into account in our analysis. In particular, the bounds
(\ref{BRXsgamma}) for $BR(B\to X_s+\gamma)$ have a strong effect
on the angle $\delta_L$, which for $\theta_{L,R}=\pi/4$ is
restricted to the ranges $0.86<\delta_L<1.35$ and
$4.93<\delta_L<5.43$.

\begin{figure}
\begin{center} \psfrag{F}{\hspace{-1cm}
$\begin{array}{c} \mathcal{A}^u~{\scriptstyle \rm (10^{-8}\
GeV)}\\ \ \end{array}$} \psfrag{m}{$\begin{array}{c} \ \\
m_{\tilde{u}}~{\scriptstyle \rm (GeV)}\end{array}$}
\psfrag{aa}{$\hspace{-.5cm} \scriptscriptstyle
m_{\tilde{s}}=1000~{\rm GeV}$} \psfrag{bb}{$\hspace{-0.5cm}
\scriptscriptstyle m_{\tilde{s}}=750~{\rm GeV}$} \psfrag{cc}{$\!\!
\scriptscriptstyle m_{\tilde{s}}=500~{\rm GeV}$}
\includegraphics[width=11cm]{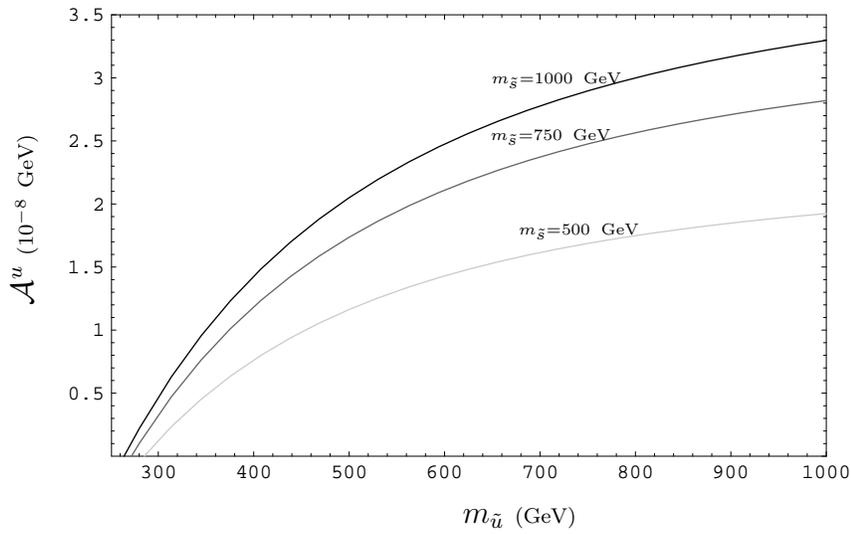}
\end{center}
\vspace{-.5cm}
\caption{\small $\mathcal{A}^u$ versus the common mass
$m_{\tilde{u}_L}=m_{\tilde{u}_R}$, for $m_{\tilde{d}_{L,R}}=
m_{\tilde{b}_{L,R}}=m_{\tilde{g}}=250~{\rm GeV}$ and
several values of $m_{\tilde{s}_{L,R}}$, in the case of
maximal $\tilde{s}-\tilde{b}$ mixing ($\theta_{R,L}=\pi/4$) and
\mbox{$\delta_L-\delta_R=\pi$}.}
\label{Au}
\end{figure}

The NP amplitude $\mathcal{A}^u$ is found to be small (even zero) for
small $u$-squark masses ($m_{\tilde{u}}\sim 250~{\rm GeV}$).  However,
for $u$- and $s$-squark masses close to 1 TeV and large
$\tilde{s}-\tilde{b}$ mixing, $\mathcal{A}^u$ can be as large as
$3.3\times 10^{-8}~{\rm GeV}$ (see figure \ref{Au}). This number
should be compared to the magnitude of the SM penguin amplitude
$|P'|\sim 3\times 10^{-8}~{\rm GeV}$. Since these values are similar,
SUSY contributions can yield important deviations from SM
predictions.

If we now allow for the variation of the SUSY parameters we find
that $A_{mix}$ can take any possible value. Correspondingly,
$A_{dir}$ can take any positive value, and negative values down to $-0.5$.
A deviation from the SM range shown in Table 1
could be explained within SUSY.

Turning to the branching ratio, it can also receive a sizeable
correction from the gluino contribution compared to the SM
prediction. It can be almost 90\% larger than the SM prediction in
the U-spin limit. Even if one includes a large uncertainty of
$\pm$ 20\% from the U-spin breaking parameter $\xi$, the
supersymmetric prediction for the $\bskk$ branching ratio can be
up to 30\% larger than that of the SM in the same regions of the
SUSY parameter space. The same applies to the ratio  $R_{d}^{s \;
susy}$.

Table 2 summarizes all of these results. We see that there is a wide
range in the values of the observables which are not allowed by SM but
that are easily accommodated by minimal SUSY. We stress that this
result holds even in a situation of quite constrained parameter space,
large hadronic uncertainties and a $\pm$20\% of SU(3) breaking in
$\xi$.

\begin{table}
\begin{center}
\begin{tabular}{|c|c|c|c|} \hline $BR_{\sss KK}^{susy}\ (\times
10^6)$ & $R_{d}^{s \; susy}$ &
$A_{dir\ \sss KK}^{susy}$ & $A_{mix\ \sss KK}^{susy}$ \\
\hline
$(3.6,79.1)$ & (0.7,17.2) & $(-0.5,1.0)$ & $(-1,1)$ \\
\hline \end{tabular}
\end{center}
\caption{\small
Allowed ranges of the $\bskk$ observables, including both SM $+$ SUSY
contributions.}
\end{table}

\section{Conclusions}

At present, there are several hints of new physics (NP) in processes
governed by $\btos$ transitions. For this reason, it is useful to
consider the effect of NP on $\btos$ processes. One such decay is
$\bskk$. There are many possible NP contributions to $\bskk$ decays.
However, to a good approximation, all of these have strong phases
which are small compared to those of the standard model (SM), and can
therefore be neglected. In this limit, one can combine all NP
contributions into a single term, parametrized by its magnitude
$\ANPu$ and weak phase $\Phi_u$.

In this paper, we have calculated the main supersymmetric (SUSY)
contributions to $\ANPu$ and $\Phi_u$, assuming that $\bdpipi$ is
unaffected.  There are many SUSY effects.  However, the principal ones
come from squark-gluino loops, which involve strong couplings, and
because $(\alpha_s/\alpha)(\mw^2/M_{\sss NP}^2) \sim 1$, they are not
suppressed compared to the SM ($M_{\sss NP} \sim 1$ TeV).  These are
expected to be the dominant effects, and so we have included only
these contributions. We have used naive factorization to compute the
matrix elements and used data from $B\to\pi\pi$ decays to estimate the
SM contribution.

In the presence of such SUSY contributions, the predictions of the
SM for $\bskk$ decays can be significantly modified, particularly
for $u$- and $s$-squark masses close to 1 TeV and large
$\tilde{s}-\tilde{b}$ mixing. For example, we have found that the
branching ratio can be increased. Even if one takes into account
the large uncertainty due to the breaking of flavor SU(3)
symmetry, the prediction of the SM $+$ SUSY for the $\bskk$
branching ratio can be up to 30\% larger than that of the SM
alone.

The situation is even more dramatic for the CP-violating
asymmetries $A_{dir}$ and $A_{mix}$. In the SM, these are
predicted to be small, with $A_{dir}$ taking positive values only.
On the other hand, in the presence of SUSY contributions, the
range of $A_{mix}$ gets enlarged from $-1$ to $1$, and $A_{dir}$
covers all the positive range, and also admits negative values
forbidden to the SM.

We therefore conclude that the study of $\bskk$ decays is very useful
with respect to new physics. The measurement of its observables can be
used to detect the presence of NP. Furthermore, the precise values of
these quantities can be used to constrain the NP parameter space. In
particular, this holds true for the case of SUSY, which can
significantly modify the SM predictions.

\bigskip
\noindent
{\bf Acknowledgements}: \\
We thank Kang Young Lee for helpful discussions. J.M. thanks A. Masiero
for useful comments. This work was
financially supported by NSERC of Canada (SB \& DL), and by
FPA2002-00748 (JM \& JV) and the Ramon y Cajal Program (JM).\\\\

\section*{Appendix A: Wilson Coefficients}

Using the approximations discussed in the text, the non-vanishing
Wilson coefficients (WC's) at the heavy scale ($\mw$) are given by
\bea c_1^q & = &
\frac{\alpha_s^2\sin{2\theta_\lft}e^{i\delta_\lft}}{4\sqrt{2}\gf
m_{\tilde{g}}^2} \left[
\frac{1}{18}F(x_{\tilde{b}_\lft\tilde{g}},x_{\tilde{q}_\rht\tilde{g}})-
\frac{5}{18}G(x_{\tilde{b}_\lft\tilde{g}},x_{\tilde{q}_\rht\tilde{g}})+
\frac{1}{2}A(x_{\tilde{b}_\lft\tilde{g}})+\frac{2}{9}B(x_{\tilde{b}_\lft\tilde{g}})
\right] \nn\\
& & \hskip1truecm -(x_{\tilde{b}_\lft\tilde{g}} \to x_{\tilde{s}_\lft\tilde{g}}) \nn\\
c_2^q & = &
\frac{\alpha_s^2\sin{2\theta_\lft}e^{i\delta_\lft}}{4\sqrt{2}\gf
m_{\tilde{g}}^2} \left[
\frac{7}{6}F(x_{\tilde{b}_\lft\tilde{g}},x_{\tilde{q}_\rht\tilde{g}})+
\frac{1}{6}G(x_{\tilde{b}_\lft\tilde{g}},x_{\tilde{q}_\rht\tilde{g}})
-\frac{3}{2}A(x_{\tilde{b}_\lft\tilde{g}})-\frac{2}{3}B(x_{\tilde{b}_\lft\tilde{g}})
\right]\nn\\
& & \hskip1truecm -(x_{\tilde{b}_\lft\tilde{g}} \to
x_{\tilde{s}_\lft\tilde{g}}) \nn\\
c_3^q & = &
\frac{\alpha_s^2\sin{2\theta_\lft}e^{i\delta_\lft}}{4\sqrt{2}\gf
m_{\tilde{g}}^2} \left[
-\frac{5}{9}F(x_{\tilde{b}_\lft\tilde{g}},x_{\tilde{q}_\lft\tilde{g}})+
\frac{1}{36}G(x_{\tilde{b}_\lft\tilde{g}},x_{\tilde{q}_\lft\tilde{g}})+
\frac{1}{2}A(x_{\tilde{b}_\lft\tilde{g}})+\frac{2}{9}B(x_{\tilde{b}_\lft\tilde{g}})
\right]\nn\\
& & \hskip1truecm -(x_{\tilde{b}_\lft\tilde{g}} \to
x_{\tilde{s}_\lft\tilde{g}}) \nn\\
c_4^q & = &
\frac{\alpha_s^2\sin{2\theta_\lft}e^{i\delta_\lft}}{4\sqrt{2}\gf
m_{\tilde{g}}^2} \left[
\frac{1}{3}F(x_{\tilde{b}_\lft\tilde{g}},x_{\tilde{q}_\lft\tilde{g}})+
\frac{7}{12}G(x_{\tilde{b}_\lft\tilde{g}},x_{\tilde{q}_\lft\tilde{g}})-
\frac{3}{2}A(x_{\tilde{b}_\lft\tilde{g}})-\frac{2}{3}B(x_{\tilde{b}_\lft\tilde{g}})
\right] \nn\\
& & \hskip1truecm -(x_{\tilde{b}_\lft\tilde{g}} \to
x_{\tilde{s}_\lft\tilde{g}}) ~, \label{Wilson coeffs} \eea
where the functions $F$, $G$, $A$ and $B$ are
\begin{eqnarray}
F(x,y)&=&-\frac{x\ln{x}}{(x-y)(x-1)^2}-\frac{y\ln{y}}{(y-x)(y-1)^2}-\frac{1}{(x-1)(y-1)}\nn\\
G(x,y)&=&\frac{x^2\ln{x}}{(x-y)(x-1)^2}+\frac{y^2\ln{y}}{(y-x)(y-1)^2}+\frac{1}{(x-1)(y-1)}\nn\\
A(x)&=&\frac{1}{2(1-x)}+\frac{(1+2x)\ln{x}}{6(1-x)^2}\nn\\
B(x)&=&-\frac{11-7x+2x^2}{18(1-x)^3}-\frac{\ln{x}}{3(1-x)^4} ~,
\label{eq:loop_fn}
\end{eqnarray}
and $x_{\tilde{q}_i\tilde{g}}\equiv
m_{\tilde{q}_i}^2/m_{\tilde{g}}^2$, where $m_{\tilde{q}_i}$ ($q=d,u$)
is the mass of the $i^{th}$ squark mass eigenstate. The expressions
for the coefficients $\tilde{c}_i^q$ are obtained from those in
Eq.~(\ref{Wilson coeffs}) via the exchange $L\leftrightarrow R$. Note
that there is a relative sign difference between our $c_4^q$ and that
given in Ref.~\cite{trojan}. Our computation of the WC's agrees with
Ref.\cite{KN}.

Using the same approximations, the SUSY contribution to the WC of
the chromomagnetic operator is given by
\beq \lambda_t \frac{2 \alpha_s}{3 \pi} C^{\rm SUSY}_{8g} =
\frac{8}{3} \frac{\alpha_s^2 \sin 2\theta_L e^{i \delta_L}}{4
\sqrt{2} \gf m_{\tilde{g}}^2} \left[f_8^{\rm SUSY}(x_{\tilde{b}_L
\tilde{g}})-(b_L \leftrightarrow s_L)\right] ~, \eeq
where the loop function is
\beq f_8^{\rm SUSY}(x)= \frac{-11 + 51\,x - 21\,x^2 - 19\,x^3 +
  6\,x\,\left( -1 + 9\,x \right) \,\log (x)}{72\, {\left( -1 + x
  \right) }^4} ~.
\eeq

\section*{Appendix B: Renormalization-group evolution of the Wilson
coefficients}

The QCD evolution of the Wilson coefficients (WC's) is given by
\cite{BBL}
\beq \vec{C}(\mu)=U_5(\mu,\mw)\vec{C}(\mw) \label{RG} \eeq
where $U_5(\mu,\mw)$ is the evolution matrix. Following GNK we work at
leading order (LO)\footnote{We have evaluated the inclusion of
next-to-leading-order corrections in $\alpha_s$ in the anomalous
dimension matrix. The impact is less than 10\%.}  neglecting
electromagnetic corrections to the anomalous dimension matrix of the
operators. The chromomagnetic operator $Q_{8g}$ is included.  The
leading logarithmic approximation depends only on the leading-order
anomalous dimension matrix $\gamma^{(0)}$ \cite{BBL}. We perform the
running translating the basis to the 12 $\Delta B=1$ SM operators.
\beq H_{\rm eff}^{\sss SM} = \frac{\gf}{\sqrt{2}} \left\{ \lambda_u
\sum_{i=1}^2 C_1(\mu)Q_i^u - \lambda_t \left[\sum_{i=3}^{10}
C_i(\mu)\,Q_i + C_{7\gamma}(\mu)\,Q_{7\gamma}+
C_{8g}(\mu)\,Q_{8g} \right] \right\} ~, \eeq
where $\lambda_i \equiv V_{ib}^* V_{is}$. In the above, $Q_1^u$ and
$Q_2^u$ are tree operators, $Q_{3-6}$ are QCD penguin operators, and
$Q_{7-10}$ are electroweak penguin operators:
\beq \begin{array}{ll} Q_1^u=(\bar{b}_\alpha
u_\beta)_{V-A}(\bar{u}_\beta s_\alpha)_{V-A} &
Q_2^u=(\bar{b}_\alpha u_\alpha)_{V-A}(\bar{u}_\beta s_\beta)_{V-A}
\\
Q_3=(\bar{b}_\alpha s_\alpha)_{V-A} \sum_q (\bar{q}_\beta
q_\beta)_{V-A} & Q_4=(\bar{b}_\alpha s_\beta)_{V-A} \sum_q
(\bar{q}_\beta
q_\alpha)_{V-A} \\
Q_5=(\bar{b}_\alpha s_\alpha)_{V-A} \sum_q (\bar{q}_\beta
q_\beta)_{V+A} & Q_6=(\bar{b}_\alpha s_\beta)_{V-A} \sum_q
(\bar{q}_\beta
q_\alpha)_{V+A} \\
Q_7= {3\over 2}(\bar{b}_\alpha s_\alpha)_{V-A} \sum_q e_q
(\bar{q}_\beta q_\beta)_{V+A} & Q_8= {3\over 2} (\bar{b}_\alpha
s_\beta)_{V-A} \sum_q e_q
(\bar{q}_\beta q_\alpha)_{V+A} \\
Q_9= {3\over 2}(\bar{b}_\alpha s_\alpha)_{V-A} \sum_q e_q
(\bar{q}_\beta q_\beta)_{V-A} & Q_{10}= {3\over 2} (\bar{b}_\alpha
s_\beta)_{V-A} \sum_q e_q
(\bar{q}_\beta q_\alpha)_{V-A} \\
Q_{7\gamma}= (e / 8\pi^2) m_b {\bar b} \sigma_{\mu \nu} (1 -
\gamma_5) F_{\mu \nu} s & Q_{8g}= (g_s / 8\pi^2) m_b {\bar b}
\sigma_{\mu \nu} (1 - \gamma_5) G_{\mu \nu} s ~, \end{array} \eeq
where $e_q$ is the electric charge of quark $q$.

In this basis the evolution matrix is $12 \times 12$, with the
coefficients $C_i$ related to the NP $c_i^q$'s in Eq.~(\ref{H})
through
\beq \begin{array}{ll} c_1^u=-\lambda_t \,(C_5 + C_7)\qquad\quad &
c_1^d=-\lambda_t\,(C_5 - \frac{1}{2}C_7)\\
c_2^u=-\lambda_t\,(C_6+C_8)\qquad &
c_2^d=-\lambda_t\,(C_6 -\frac{1}{2}C_8)\\
c_3^u=-\lambda_t\,(C_3+C_9)\qquad &
c_3^d=-\lambda_t\,(C_3 -\frac{1}{2}C_9)\\
c_4^u=-\lambda_t\,(C_4 + C_{10})\qquad & c_4^d=-\lambda_t\,(C_4
-\frac{1}{2}C_{10}) \label{c's vs C's} \end{array} \eeq
Note that the $c_{5,6}^{u,d}$ are zero at the $\mw$ scale in our
case. We take them to be zero also at the $m_b$ scale, since the
electroweak combination $c_{5,6}^u(m_b)-c_{5,6}^d(m_b)$ is at LO a
function only of $c_{5,6}^{u,d}(\mw)$, and the QCD combination
$(c_{5,6}^{u}(m_b)+2 c_{5,6}^d(m_b))/3$ is mostly dominated by the
same combination at $\mw$, taking into account that all NP penguin
WC's are of similar size.



\begin{thebibliography}{99}

\bibitem{sin2betapeng} K.~F.~Chen [Belle Collaboration],
[hep-ex/0504023]; B.~Aubert {\it et al.}  [BABAR Collaboration],
[hep-ex/0503011]; B.~Aubert {\it et al.}  [BABAR Collaboration],
[hep-ex/0502019]; B.~Aubert {\it et al.}  [BABAR Collaboration],
[hep-ex/0502017].

\bibitem{BVVTP} For a study of triple products in the SM and with new
physics, see A.~Datta and D.~London, Int.\ J.\ Mod.\ Phys.\ A {\bf
19}, 2505 (2004).

\bibitem{phiKstarTP} B.~Aubert {\it et al.}  [BABAR
Collaboration], Phys.\ Rev.\ Lett.\ {\bf 93}, 231804 (2004);
K.~Senyo [Belle Collaboration], hep-ex/0505067.

\bibitem{BphiKstar_exp} B.~Aubert {\it et al.}  [BABAR
Collaboration], Phys.\ Rev.\ Lett.\ {\bf 91}, 171802 (2003);
K.~F.~Chen {\it et al.} [BELLE Collaboration], hep-ex/0503013.

\bibitem{BphiKstarNP} C.~Dariescu, M.~A.~Dariescu, N.~G.~Deshpande and
D.~K.~Ghosh, Phys.\ Rev.\ D {\bf 69}, 112003 (2004); E.~Alvarez,
L.~N.~Epele, D.~G.~Dumm and A.~Szynkman, hep-ph/0410096; Y.~D.~Yang,
R.~M.~Wang and G.~R.~Lu, Phys.\ Rev.\ D {\bf 72}, 015009 (2005);
P.~K.~Das and K.~C.~Yang, Phys.\ Rev.\ D {\bf 71}, 094002 (2005);
C.~S.~Kim and Y.~D.~Yang, hep-ph/0412364; C.~H.~Chen and C.~Q.~Geng,
Phys.\ Rev.\ D {\bf 71}, 115004 (2005); C.~S.~Huang, P.~Ko, X.~H.~Wu
and Y.~D.~Yang, arXiv:hep-ph/0511129.

\bibitem{BphiKstarSM} P.~Colangelo, F.~De Fazio and T.~N.~Pham,
Phys.\ Lett.\ B {\bf 597}, 291 (2004); M.~Ladisa, V.~Laporta,
G.~Nardulli and P.~Santorelli, Phys.\ Rev.\ D {\bf 70}, 114025
(2004); A.~L.~Kagan, Phys.\ Lett.\ B {\bf 601}, 151 (2004);
W.~S.~Hou and M.~Nagashima,hep-ph/0408007; H.~Y.~Cheng, C.~K.~Chua
and A.~Soni,Phys.\ Rev.\ D {\bf 71}, 014030 (2005).

\bibitem{BrhoKstar_exp} Particle Data Group, Ref. \cite{pdg};
B.~Aubert [BABAR Collaboration], hep-ex/0408093; J.~Zhang {\it et
al.}  [BELLE Collaboration], hep-ex/0505039.

\bibitem{BrhoKstar} S.~Baek, A.~Datta, P.~Hamel, O.~F.~Hernandez
and D.~London, hep-ph/0508149.

\bibitem{BKpiexp} CLEO Collaboration, S.~Chen {\it et al.}, Phys.\
Rev.\ Lett.\ {\bf 85}, 525 (2000); CLEO Collaboration, A.~Bornheim
{\it et al.}, Phys.\ Rev.\ D {\bf 68}, 052002 (2003); Belle
Collaboration, Y.~Chao {\it et al.}, Phys.\ Rev.\ D {\bf 69}, 111102
(2004); hep-ex/0407025, Phys.\ Rev.\ Lett.\ {\bf 93}, 191802 (2004);
BELLE Collaboration, K.~Abe {\it et al.}, hep-ex/0409049; BABAR
Collaboration, B.~Aubert {\it et al.}, Phys.\ Rev.\ Lett.\ {\bf 89},
281802 (2002), hep-ex/0408062, hep-ex/0408080, hep-ex/0408081, Phys.\
Rev.\ Lett.\ {\bf 93}, 131801 (2004).

\bibitem{BKpidecays0} R.~Fleischer and T.~Mannel, Phys.\ Rev.\ D {\bf
57}, 2752 (1998), ibid, hep-ph/9706261; A.~J.~Buras, R.~Fleischer and
T.~Mannel, Nucl.\ Phys.\ B {\bf 533}, 3 (1998); R.~Fleischer, Phys.\
Lett.\ B {\bf 435}, 221 (1998); M.~Neubert and J.~L.~Rosner, Phys.\
Rev.\ Lett.\ {\bf 81}, 5076 (1998); M.~Neubert and J.~L.~Rosner,
Phys.\ Lett.\ B {\bf 441}, 403 (1998); A.~J.~Buras and R.~Fleischer,
Eur.\ Phys.\ J.\ C {\bf 11}, 93 (1999).

\bibitem{BKpidecays1} M.~Beneke, G.~Buchalla, M.~Neubert and
C.~T.~Sachrajda, Nucl.\ Phys.\ B {\bf 606}, 245 (2001); M.~Beneke and
M.~Neubert, Nucl.\ Phys.\ B {\bf 675}, 333 (2003).

\bibitem{BKpidecays2} R.~Fleischer and J.~Matias, Phys.\ Rev.\ D {\bf
61}, 074004 (2000), Phys.\ Rev.\ D {\bf 66}, 054009 (2002).

\bibitem{BKpidecays2bis} J.~Matias, Phys.\ Lett.\ B {\bf 520}, 131
(2001).

\bibitem{BKpidecays3} A.~J.~Buras, R.~Fleischer, S.~Recksiegel and
F.~Schwab, Eur.\ Phys.\ J.\ C {\bf 32}, 45 (2003), Phys.\ Rev.\
Lett.\ {\bf 92}, 101804 (2004), Nucl.\ Phys.\ B {\bf 697}, 133
(2004), Acta Phys.\ Polon.\ B {\bf 36}, 2015 (2005).

\bibitem{BKpidecays4} V.~Barger, C.~W.~Chiang, P.~Langacker and
 H.~S.~Lee, Phys.\ Lett.\ B {\bf 598}, 218 (2004); S.~Mishima and
 T.~Yoshikawa, Phys.\ Rev.\ D {\bf 70}, 094024 (2004); Y.~L.~Wu and
 Y.~F.~Zhou, Phys.\ Rev.\ D {\bf 71}, 021701 (2005); H.~Y.~Cheng,
 C.~K.~Chua and A.~Soni, Phys.\ Rev.\ D {\bf 71}, 014030 (2005);
 Y.~Y.~Charng and H.~n.~Li, Phys.\ Rev.\ D {\bf 71}, 014036 (2005);
 X.~G.~He and B.~H.~J.~McKellar, hep-ph/0410098; S.~Baek, P.~Hamel,
 D.~London, A.~Datta and D.~A.~Suprun, Phys.\ Rev.\ D {\bf 71}, 057502
 (2005); C.~S.~Kim, S.~Oh and C.~Yu, hep-ph/0505060; H.~n.~Li,
 S.~Mishima and A.~I.~Sanda, hep-ph/0508041.

\bibitem{GHLR} M.~Gronau, O.~F.~Hern\'andez, D.~London and
J.~L.~Rosner, Phys.\ Rev.\ D {\bf 50}, 4529 (1994), Phys.\ Rev.\ D
{\bf 52}, 6374 (1995).

\bibitem{DL} A.~Datta and D.~London,
Phys.\ Lett.\ B {\bf 595}, 453 (2004).

\bibitem{LMV} D.~London, J.~Matias and J.~Virto, Phys.\ Rev.\ D {\bf
71}, 014024 (2005).

\bibitem{Wolfenstein} L.~Wolfenstein, Phys.\ Rev.\ Lett.\ {\bf 51},
  1945 (1983).

\bibitem{CKMfitter} The CKMfitter group, http://www.slac.stanford.edu \\
\null~~~~/xorg/ckmfitter/ckm\_results\_winter2005.html.

\bibitem{Chen:2001sx} C.~H.~Chen, Phys.\ Lett.\ B {\bf 520}, 33
  (2001).

\bibitem{Fleischer:1999} R.~Fleischer, Phys.\ Lett.\ B {\bf 459}, 306
  (1999)

\bibitem{LonMat} D.~London and J.~Matias, Phys.\ Rev.\ D {\bf 70},
  031502 (2004).

\bibitem{Mannel} A.~Khodjamirian, T.~Mannel and M.~Melcher, Phys.\
Rev.\ D {\bf 68}, 114007 (2003).

\bibitem{Aubert:2005ni} B.~Aubert {\it et al.}  [BABAR Collaboration],
  hep-ex/0508046.

\bibitem{Chao:2003ue} Y.~Chao {\it et al.}  [Belle Collaboration],
  Phys.\ Rev.\ D {\bf 69}, 111102 (2004).

\bibitem{Aubert:2005av} B.~Aubert {\it et al.}  [BaBar Collaboration],
  hep-ex/0501071.

\bibitem{AbeLP05} Belle Collaboration, talk by K.Abe at the
  Lepton-Photon conference, 2005.

\bibitem{Abe:2005dz} K.~Abe {\it et al.}  [Belle Collaboration],
  hep-ex/0502035.


\bibitem{trojan} Y.~Grossman, M.~Neubert and A.~L.~Kagan, JHEP {\bf
9910}, 029 (1999).

\bibitem{pdg} Particle Data Group Collaboration, S.~Eidelman {\it et
al.}, Phys.\ Lett.\ B {\bf 592} (2004) 1.

\bibitem{HFAG} The Heavy Flavor Averaging Group,
http://www.slac.stanford.edu/xorg/hfag/

\bibitem{Alexander:2005cx} J.~Alexander {\it et al.}  [Heavy Flavor
  Averaging Group (HFAG)], hep-ex/0412073.

\bibitem{KN} A.~L.~Kagan and M.~Neubert, Phys.\ Rev.\ Lett.\ {\bf 83},
4929 (1999).

\bibitem{BBL} A.~J.~Buras, M.~Jamin and M.~E.~Lautenbacher, Nucl.\
 Phys.\ B {\bf 408}, 209 (1993); G.~Buchalla, A.~J.~Buras and
 M.~E.~Lautenbacher, Rev.\ Mod.\ Phys.\ {\bf 68}, 1125 (1996).

\end{thebibliography}
\end{document}